\documentclass{jpsj3}  
\usepackage{txfonts} 
\usepackage[dvipdfmx]{graphicx}

\title{Phase Separation Induced by Density-Dependent Hopping Terms
}
 
\author{Kazuhiro Kuboki \thanks{kuboki@kobe-u.ac.jp}}
\inst{Department of Physics, Kobe University, Kobe 657-8501, Japan
}  
 
\abst{
We study phase separation in the $t-t'-J$ model 
in which the constraint of no double occupancy  
due to strong electron correlations leads to 
effective narrowing of the band width.
Using a mean-field approximation, we calculate the compressibility in 
the normal state and show that the phase separation can be 
induced by the variation of the band width.
}
 
\begin{document}

\maketitle

\newpage

Spontaneous lattice symmetry breaking in electron density 
is widely observed in strongly correlated electron systems, e.g., 
high-$T_c$ cuprate superconductors in which charge ordered states
occur in a wide region of the temperature ($T$)- 
doping rate ($\delta$) phase diagram.\cite{PSREV}
In theoretical models without long-range Coulomb interactions such as 
the Hubbard and $t-J$ models, phase separation (PS) is known to 
appear as a typical inhomogeneous electron density 
state.\cite{Emery,Putikka,Dagotto,Yokoyama, Kohno,Hellberg,Shih,
Calandra,White,Ivanov,Macridin,Aichhorn,Aichhorn2,Igoshev,
Khatami,Otsuki,Bejas,Bejas2,OF} 
Especially in Refs. 12 and 13, the location of the PS  
region in the phase diagram was intensively investigated by
applying the $1/N$ expansion method to the $t-J$ model.  

In the $t-J$ model, double occupancy of a single site is 
prohibited due to the strong electron correlation. 
When this condition is treated using slave-boson mean-field  
 (SBMF) or Gutzwiller approximations, 
transfer integrals are multiplied by the doping rate $\delta$ $(<1)$ 
and thus reduced. These density-dependent transfer integrals lead to   
the narrowing of the band width.
Recently, it is demonstrated that density-dependent hopping 
terms exist in the effective single-band Hamiltonian for the cuprates 
that is derived from the three band model using the density matrix 
renormalization group method.\cite{Jiang2} 
This suggests that the hopping process in strongly correlated 
electron systems may depend on the site occupancy in general, and 
it may lead to spontaneous lattice symmetry breaking of electron density, 
namely, phase separation and charge order.

In this short note, we study the effect of density-dependent 
hopping terms on the phase separation. 
We employ the $t-t'-J$ model ($t-J$ model with extended transfer
integrals) and treat it within the SBMF approximation 
as a test case to explore this problem. 
We investigate the compressibility in the normal metallic 
(uniform RVB) state but not in ordered states,
in order  to focus on the band narrowing effect on the phase separation.

We treat the $t-t'-J$ model on a square lattice with the 
Hamiltonian, 
\begin{eqnarray}
\displaystyle 
H = \displaystyle -\sum_{j,\ell,\sigma} 
t_{j\ell} {\tilde c}^\dagger_{j\sigma} {\tilde c}_{\ell\sigma}
 +J\sum_{\langle j,\ell\rangle} {\bf S}_j\cdot {\bf S}_\ell,   
\end{eqnarray}
where the transfer integrals $t_{j\ell}$ are finite for the 
first-  ($t$) and second-  ($t'$) nearest-neighbor bonds, 
or zero otherwise. 
$J$ is the antiferromagnetic superexchange
interaction, and $\langle j,\ell \rangle$ denotes nearest-neighbor bonds.
${\tilde c}_{j\sigma}$ is the operator for the electrons 
in Fock space without double occupancy. 
We treat this condition using the SBMF 
theory assuming the Bose condensatoin of holons,\cite{OF,LNW}
and write spinon operator as $f_{j\sigma}$.

In order to study the phase separation in the normal state, 
we decouple this Hamiltonian by employing only 
$\chi_f = \sum_\sigma \langle f^\dagger_{j\sigma}f_{j+\eta\sigma}\rangle$  
($\eta = \pm {\hat x}, \pm {\hat y}$) as an order parameter (bond order parameter). 
Selfconsistency equations are obtained by minimizing the free 
energy and given as, 
\begin{eqnarray}
\displaystyle n = \frac{2}{N}\sum_k f(\xi_k), \ \ 
\chi_ f  = \frac{1}{N}\sum_k \gamma_k  f(\xi_k)
\end{eqnarray}
where $n$ ($=1-\delta)$ and $N$ are the electron density and 
the total number of lattice sites, respectively, 
$\gamma_k = \cos k_x + \cos k_y $ and 
$\xi_k = -2(t\delta + \frac{3J}{8}\chi_f)\ \gamma_k 
-4t'\delta\cos k_x\cos k_y - \mu$ with $\mu$ being the chemical potential. 
$f(\xi_k)$ is the Fermi distribution function.

By taking derivatives of Eq.(2) with respect to $\mu$,  
we obtain simultaneous equations for the compressibility,  
$\partial n/\partial \mu$, and 
$\partial \chi_f/\partial \mu$,  
\begin{eqnarray}
\displaystyle {\hat A}
\left (\begin{array}{cc} 
\displaystyle 
\frac{\partial n}{\partial \mu} \\
\displaystyle
\frac{\partial \chi_f}{\partial \mu}
\end{array}\right )
= 
\left (\begin{array}{cc}
c_1 \\ 
c_2
\end{array}\right ),   
\end{eqnarray}
where ${\hat A}$ is a $2\times 2$ matrix with 
$A_{11} = 1 - \frac{2}{N}\sum_k \Gamma_k f'(\xi_k)$, 
$A_{12} = \frac{3J}{2N}\sum_k \gamma_k f'(\xi_k)$, 
$A_{21} = - \frac{1}{N}\sum_k \gamma_k\Gamma_k f'(\xi_k)$, and 
$A_{22} = 1 + \frac{3J}{4N}\sum_k \gamma_k^2 f'(\xi_k)$.
Here $\Gamma_k = 2t\gamma_k + 4t'\cos k_x\cos k_y$,  
$c_1 = -\frac{2}{N}\sum_k f'(\xi_k)$, and 
$c_2 = -\frac{1}{N}\sum_k \gamma_k  f'(\xi_k)$. 
By solving Eq.(3), we get 
$
\partial n/\partial \mu
=   (A_{22}c_1 - A_{12}c_2)/D
$
and  
$
\partial \chi_f/\partial \mu 
=   -(A_{21}c_1 - A_{11}c_2)/D
$  
with $D = A_{11}A_{22}-A_{12}A_{21}$.

For comparison, we also calculate the compressibility by dropping 
the derivatives of $\delta$'s in $\xi_k$.
Namely, we treat them  constants independent of $\mu$.
It simply amounts to set $\Gamma_k=0$, and so $A_{11} =1$ and $A_{21} = 0$.
We denote the compressibility in this case as $(\partial n/\partial \mu)_0$.

In Fig.1 we show the results for the 
compressibility, $\partial n/\partial \mu$, 
for several choices of $T$. 
We take the parameters 
$J=1$, $t/J=2.5$, $t'/t= \pm 0.3$.
The $t'$ term is introduced to change the Fermi surface 
and to examine its effect.
In the case of  $t'/t = 0.3$, the compressibility is positive for 
$\delta \gtrsim 0.1$, i.e., the region relatively away from half filling, 
and the homogeneous state is stable here. 
The compressibility diverges at the transition point 
$\delta \sim 0.1$, and becomes negative near half-filling.
Thus the homogeneous state is not stable in this region and 
the phase separation occurs.
In numerical calculations, if we take $\delta$ as an input parameter
a spurious solution may be obtained, 
but the compressibility diverges (or becomes negative). 
When the resultant $\mu$ is taken as an input parameter, 
we cannot obtain the original value of $\delta$,
since multiple values of $\delta$ corresponds to a single value of $\mu$.
In the case of $t'/t = -0.3$, the compressibility is always finite 
and positive in the 
region studied and so the homogeneous state is stable. 
This means that the occurence of phase separation depends on 
the band structure, in other words, the shape of the Fermi surface. 
In Ref.13,  the PS region also appears for negative $t'$ cases. 
The reason for the difference is that the Hamiltonian in Ref.13 
includes the term $-J\sum_{\langle j,l \rangle}n_jn_l/4$, which 
is an attractive interaction between nearest-neighbor sites, and 
it will help to induce (enhance) the PS state. 
(Actually if we include this term in $H$, phase separated states 
occur for  $t'/t = -0.3$ near $\delta=0$.)

In Fig.2 we present the results for $(\partial n/\partial \mu)_0$, 
which does not take into account the variation of the band width with $\mu$.
It is seen that $(\partial n/\partial \mu)_0$ is always finite and positive 
for both $t'/t = 0.3$ and $t'/t = -0.3$, 
in contrast to the case of $\partial n/\partial \mu$. 
This indicates that the variation of the band width 
is responsible for the appearance of the phase separation.

Results for $\partial \chi_f/\partial \mu$ are shown in Fig.3 
for both $t'/t = 0.3$ and $t'/t = -0.3$. The former diverges at the 
phase transition point, while the latter is an analytic function,  
as expected from the behavior of $\partial n/\partial \mu$.

\begin{figure}[htb]
\begin{center}
\includegraphics[width=7.5cm,clip]{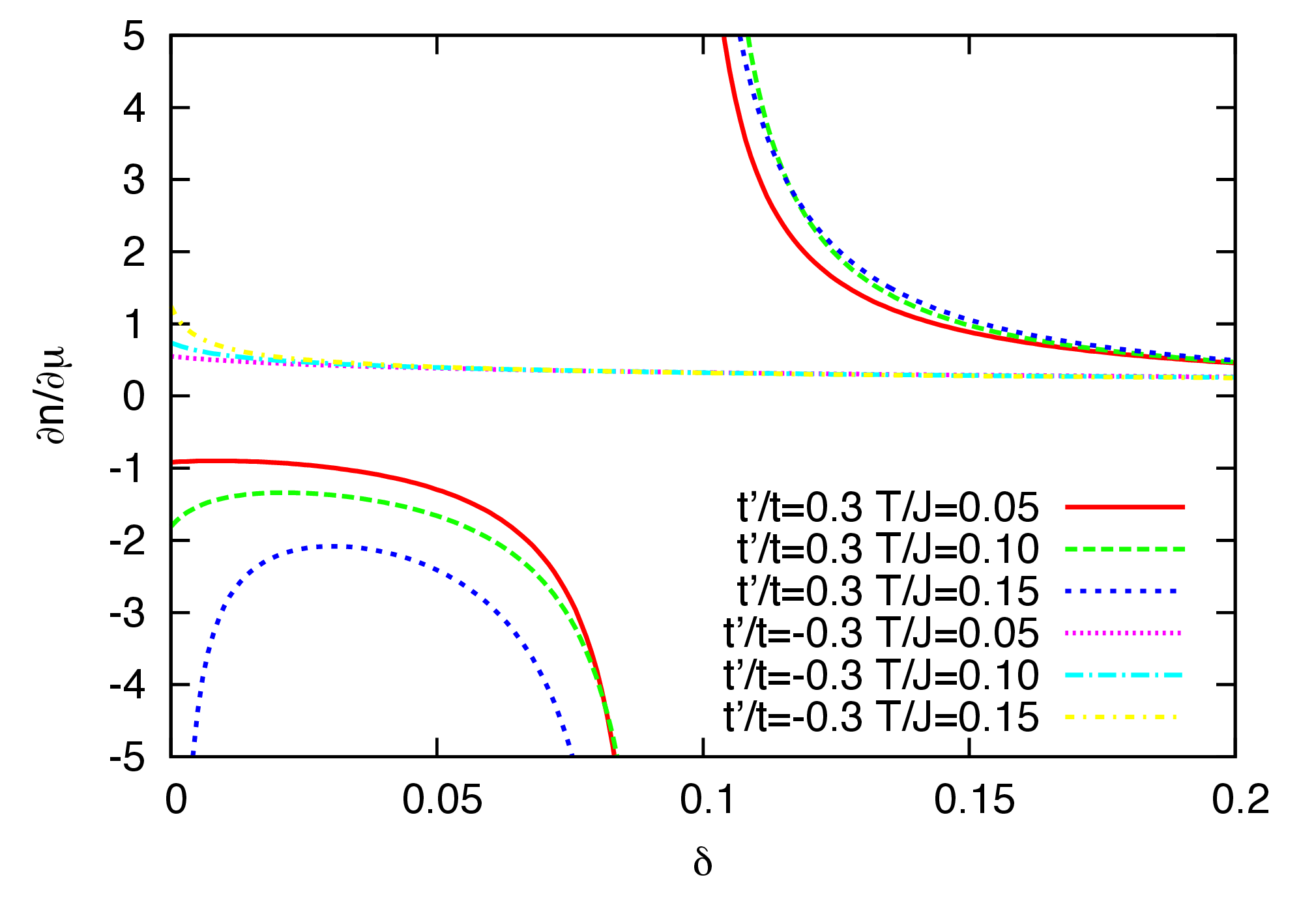}
\caption{(Color online) Compressibility $\partial n/\partial \mu$ for 
$t'/t = 0.3$ and $t'/t = -0.3$ with temperatures $T/J = 0.05$,
$T/J = \ 0.10$, and $T/J =0.15$.
} 
\end{center}
\end{figure}

\begin{figure}[htb]
\begin{center}
\includegraphics[width=7.5cm,clip]{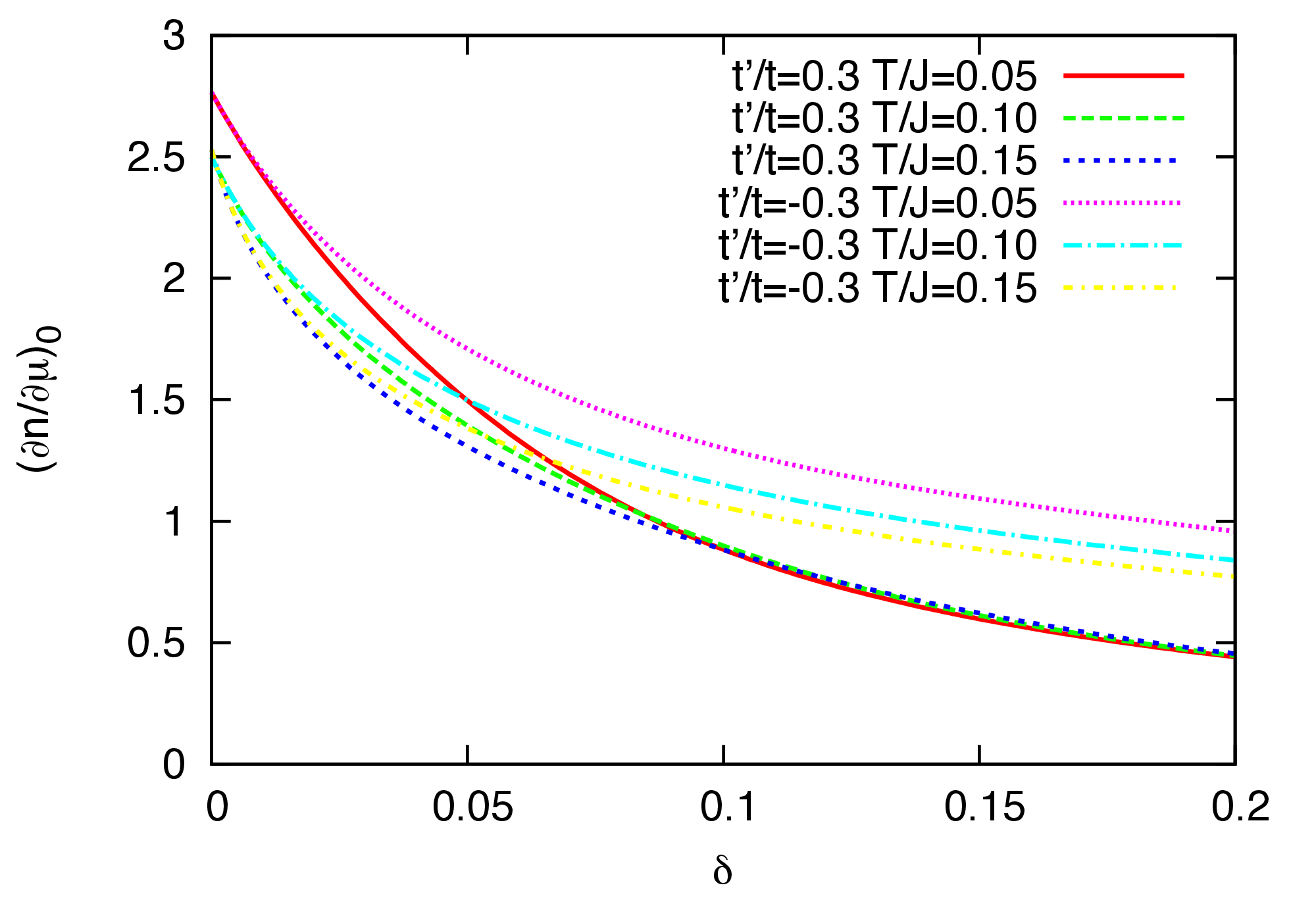}
\caption{(Color online)   Compressibility $(\partial n/\partial \mu)_0$ 
for $t'/t = 0.3$ and $t'/t = -0.3$ with temperatures $T/J = 0.05$,
$T/J = \ 0.10$, and $T/J =0.15$.
} 
\end{center}
\end{figure}

\begin{figure}[htb]
\begin{center}
\includegraphics[width=7.5cm,clip]{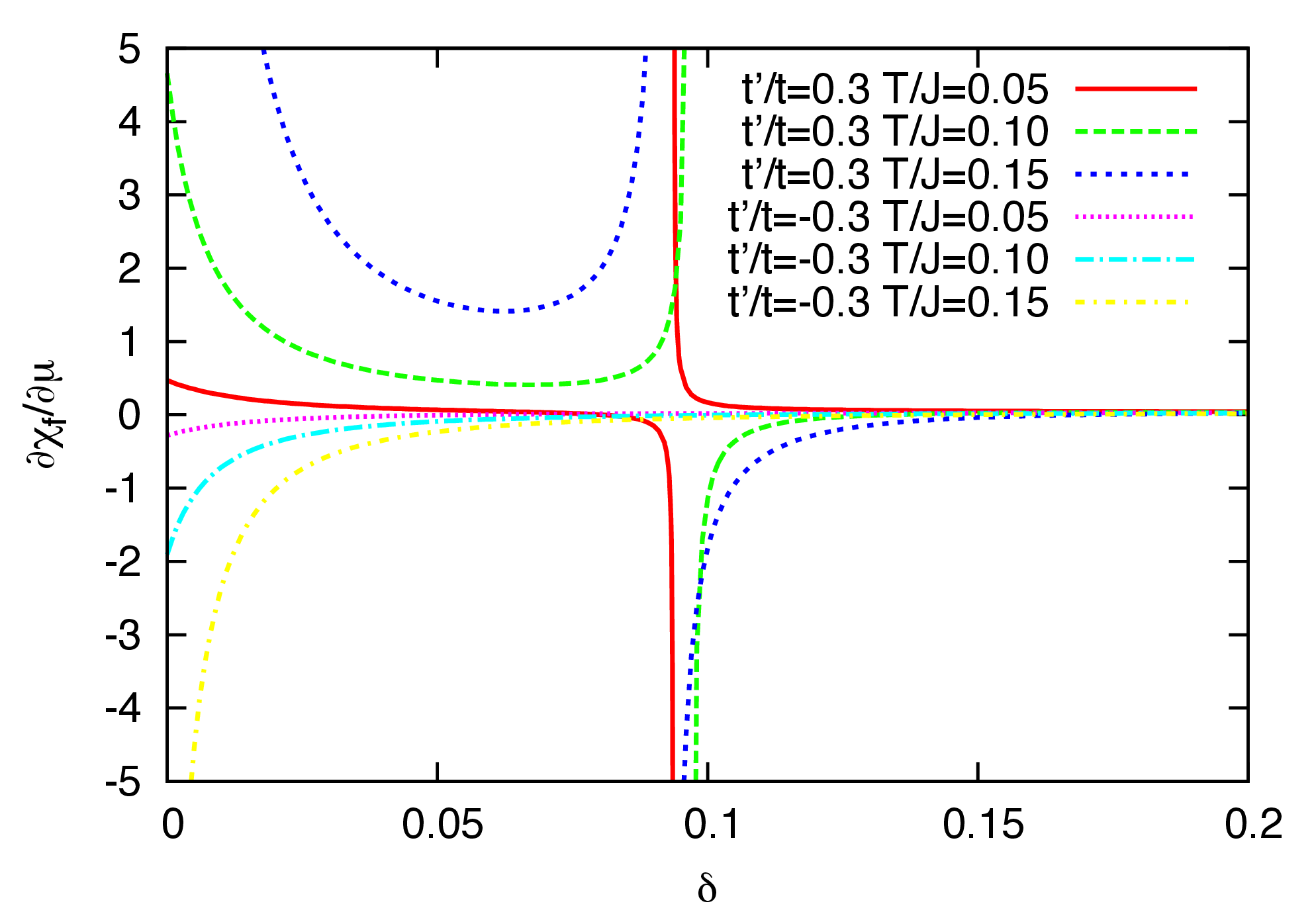}
\caption{(Color online)  $\partial \chi_f/\partial \mu$ 
for $t'/t = 0.3$ and $t'/t = -0.3$ with temperatures $T/J = 0.05$,
$T/J = \ 0.10$, and $T/J =0.15$.
} 
\end{center}
\end{figure}

In summary, we have calculated the compressibility in 
the normal state of the $t-t'-J$ model. 
It is found that the variation of the band width due to strong electron 
correlations can induce phase separation, 
but its occurence depends on the band structure, in other words, 
the shape of the Fermi surface.

The density-dependent hopping terms may induce charge ordered states 
as well as phase separation, and it will be intriguing to 
explore this possibility.
In these studies, the inclusion of the long-range Coulomb 
interaction may be necessary, since it would suppress 
the phase separated states strongly compared with other states, 
and thus affect the competition among the candidate states.

\medskip
\begin{acknowledgment}
 The author thanks M. Hayashi, H. Matsukawa, and H. Yamase for useful discussions. 
\end{acknowledgment}


\end{document}